\documentclass[twocolumn,showpacs,preprintnumbers,amsmath,amssymb]{revtex4}
\usepackage{graphicx}
\usepackage{bm}

\begin{document}
\title{Broadband THz study of excitonic resonances in the high-density regime}

\author{Rupert Huber, Robert A. Kaindl, Ben A. Schmid, and Daniel S. Chemla}
\affiliation{Department of Physics, University of California at Berkeley,
and Material Sciences Division, E.O. Lawrence Berkeley National Laboratory,
Berkeley, California 94720, USA}
\date{\today}

\begin{abstract}
We report the first terahertz study of  the intra-excitonic $1s$-$2p$ transition at high excitation densities in GaAs/AlGaAs quantum wells. A strong shift, broadening, and ultimately the disappearance of this resonance occurs with increasing density, after ultrafast photoexcitation at the near-infrared exciton line. Densities of excitons and unbound electron-hole pairs are followed quantitatively using a model of the composite terahertz dielectric response. Comparison with near-infrared absorption changes reveals a significantly enhanced energy shift and broadening of the intra-excitonic resonance.
\end{abstract}

\pacs{78.47.+p, 73.20.Mf, 78.67.De}

\maketitle
For a many-particle electron-hole ($e$-$h$) system in a photoexcited semiconductor, density-dependent Coulomb interactions determine the spectrum of its lowest-energy elementary excitations. At sufficiently low densities and temperatures, charge-neutral excitons form whose ground state excitation is the transition from $1s$ to $2p$ levels. These bound states are modified by many-body effects at high densities \cite{Ras82}. An important phenomenon in this context is the excitonic Mott transition: Driven by decreasing interparticle distance, excitonic bound states ultimately cease to exist such that a conducting plasma of unbound \mbox{$e$-$h$} pairs prevails \cite{Hau04,Kel65,Mot61,Kap05}. Quasi two-dimensional (2D) excitons in quantum wells, due to large binding energies and sharp optical resonances, are particularly suitable to study high-density many-particle effects of excitons \cite{Sha99,Che01}.

Numerous studies have investigated the near-infrared excitonic resonances just below the semiconductor band edge, and their broadening, bleaching, and energy shift due to photoexcited $e$-$h$ pairs (see e.\,g. \cite{Hon89,Tra87,Cho04,Feh82,Pey84,Wak92,Man98,Lit99,Kno86}). Broadening occurs via collisional interactions \cite{Hon89}. The energy shift is more subtle: with increasing density, phase-space filling and screening induce both a renormalization of single-particle states (and thus the band gap) as well as a reduction of the exciton binding energy \cite{Tra87,Cho04}. These two contributions counteract and cancel exactly in the three-dimensional case, where no shift of the exciton line is observed \cite{Feh82}. In quasi-2D systems, a small "blue" or "red"-shift remains depending on the conditions \cite{Pey84,Wak92,Man98,Lit99}. Hence, it is difficult to determine the density dependence of the exciton binding energy from such measurements.

Terahertz (THz) spectroscopy, in contrast, constitutes a fundamentally different approach to study many-particle states. Transitions between internal states of excitons occur in this spectral region, providing a direct measure of exciton densities and binding energies \cite{Tim76,Gro94,Cer96,Kir00,Kai03,Gal05}. A recent THz study investigated the transient conducting and insulating phases that occur upon formation and ionization of excitons \cite{Kai03}. Furthermore, THz radiation is equally sensitive to the ultrafast dynamics of many-body correlations of unbound $e$-$h$ pairs \cite{Hub01}. Thus, the interplay between optically generated excitons and unbound $e$-$h$ pairs becomes directly observable. Until now, however, THz studies investigated a dilute exciton gas, while density-dependent modifications of the pair correlations were not reported.

In this paper, we employ broadband THz spectroscopy to probe the internal 1$s$-2$p$ transition of resonantly photogenerated excitons at high densities. This technique gives direct access to the low-energy complex dielectric function of the dense electron-hole ensemble.
We observe a strong shift and broadening of the THz exciton resonance with increasing pair density. The response is reproduced by an analytical model that quantitatively determines the densities of excitons and unbound $e$-$h$ pairs. At high pump powers, excitonic resonances are absent and the response of the system is that of a conducting state. Comparing the THz spectra with the near-infrared response, we find a significantly enhanced energy shift and broadening of the intra-excitonic resonance.

\begin{figure}
\includegraphics[width=8.5cm]{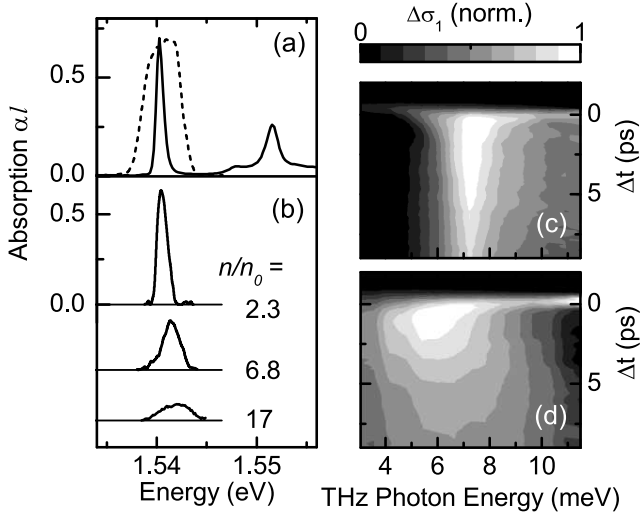}
\caption{\label{fig:epsart} (a) Solid line: near-infrared, equilibrium absorption spectrum of the quantum well sample. Dashed curve: spectrum of the pump pulses. (b) Corresponding absorption spectra at delay time $\Delta t$ = 2.5 ps after resonant $1s$HH excitation, for several densities $n$ (where $n_{0}\approx 1 \times 10^{10} \mathrm{cm}^{-2}$). The lattice temperature is $T_{\mathrm{L}}$~=~6~K. (c,d) Ultrafast dynamics of the THz conductivity $\Delta\sigma_1(\omega)$ after resonant $1s$HH excitation for initial pair densities (c) $n \approx 2 \times 10^{10} \mathrm{cm}^{-2}$
and (d) $n \approx 14 \times 10^{10} \mathrm{cm}^{-2}$.}
\end{figure}

The sample investigated here consists of a stack of ten undoped 14-nm-wide GaAs wells separated by 10-nm-wide Al$_{0.3}$Ga$_{0.7}$As barriers. All measurements are performed at low temperatures (T = 6 K). The near-infrared absorption spectrum, shown in Fig. 1(a), is dominated by the $1s$ heavy-hole (HH) exciton line at 1.540 eV. Its linewidth is 0.8 meV (full width at half maximum, FWHM), pointing to a high sample quality. With increasing energy, the $2s$HH and the $1s$ light-hole exciton appear, followed by transitions into the band-to-band continuum.

Heavy-hole excitons are excited via spectrally-shaped near-infrared pulses derived from a 250-kHz Ti:sapphire amplifier. In order to inject high pair densities, the pump spectrum [dashed line, Fig.1(a)] is tailored to a width (4 meV FWHM) distinctly larger than the low-intensity absorption line. This allows for continued overlap as the line broadens at higher densities. Figure~1(b) shows absorption spectra for several densities, monitored via near-infrared probe pulses transmitted through the sample after a delay time of $\Delta t$ = 2.5 ps. At elevated pump fluences, we observe broadening, bleaching, and a slight blue shift of the near-infrared 1$s$ absorption line. These changes are explained by the joint effects of phase-space filling and screening \cite{Tra87,Cho04,Feh82,Pey84,Wak92,Man98,Lit99}.

To probe the internal response of the bound $e$-$h$ pairs, we employ broadband THz probe pulses generated by optical rectification. The real-time evolution of the THz electric field is detected by electro-optic sampling in a 500-$\mu$m-thick ZnTe crystal. We measure both the field transmitted through the sample in equilibrium, and its pump-induced change at a fixed pump-probe delay time $\Delta t$. From this, the complex optical conductivity $\tilde \sigma(\omega)$ and its transient change is obtained. The retrieval algorithm takes into account the multi-layer quantum well structure \cite{Kai03}. In the following, we will express the full THz dielectric response as $\tilde \sigma \left( \omega  \right) = \sigma _1 \left( \omega \right) + i\omega \epsilon_0\left[ {1 - \varepsilon _1 \left( \omega  \right)}\right]$. The real part of the optical conductivity $\sigma_1(\omega)$ is a measure of the absorbed power density, while the real part of the dielectric function $\epsilon_1(\omega)$ provides a measure of the out-of-phase, inductive response.

Ultrafast transient changes in the THz conductivity $\Delta \sigma_1$ after near-infrared excitation are displayed in Figs.~1(c,d) as a function of the photon energy of the THz probe pulses and the pump-probe delay time $\Delta t$. These data, which are shown for two representative pump fluences, demonstrate a strong density dependence of the THz response. In the following, we will concentrate on features that emerge within the time resolution of our experiment of about 1~ps.

\begin{figure}
\includegraphics[width=8.5cm]{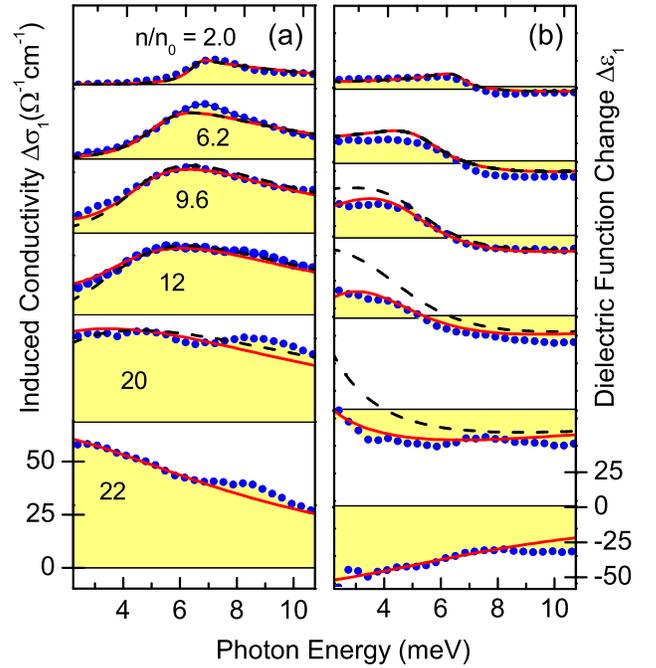}
\caption{\label{fig:epsart} (Color online). Photoinduced THz conductivity spectra $\Delta \sigma_{1}(\omega)$ [left panels] and dielectric function changes $\Delta \epsilon_{1}(\omega)$ [right panels] for various excited $e$-$h$ pair densities $n$ (given in units of $n_{0} = 10^{10} \mathrm{cm}^{-2}$). Experimental data (solid dots) are measured at pump-probe delay $\Delta t$ = 2.5 ps. Dashed lines: quasi-2D exciton model, solid lines: sum of exciton and Drude components, as explained in the text. At the two lowest densities, the free carriers vanish and the models coincide, while the model curve at the highest density shows a pure Drude response. Curves in each panel are shifted vertically for clarity, and are equally scaled.}
\end{figure}

Figure~2 shows the THz response (solid dots) for various excitation densities $n$ at delay time $\Delta t$ = 2.5 ps. For the lowest density of $n = 2 \times 10^{10}~\mathrm{cm}^{-2}$, a narrow asymmetric peak in $\Delta \sigma_1(\omega)$ at $\hbar \omega = 7\,\mathrm{meV}$ demonstrates the existence of bound {$e$-$h$} pairs \cite{Kai03}. The maximum arises from the internal transition from 1$s$ to 2$p$ bound states, while the high-energy shoulder corresponds to transitions from 1$s$ into higher bound and continuum states. The conductivity vanishes at low frequencies, which corroborates the insulating nature of this dilute exciton gas. The dispersive zero crossing of $\Delta \epsilon_1$ at a photon energy of 7\,meV is characteristic of the well-defined intra-excitonic oscillator.

With increasing excitation density, three profound changes occur in the THz response as revealed in Fig.~2: $(i)$ The area enclosed by the conductivity curve $\Delta \sigma_{1}(\omega)$ increases, while the near-infrared exciton line shown in Fig.\,1(b) bleaches. Thus, with rising excitation density, oscillator strength is increasingly transferred from interband to intraband transitions. $(ii)$ The width of the observed THz features strongly increases. $(iii)$ Both the spectral maximum of $\Delta \sigma_1(\omega)$ and the zero crossing of $\Delta \epsilon_1(\omega)$ shift to lower frequencies. At the highest excitation density, the conductivity rises monotonically towards lower frequencies and a zero crossing of the dielectric function is not observed. This behavior is indicative of a conducting phase in the absence of excitonic resonances.

For a quantitative analysis, we model the experimentally determined THz response (Fig.\,2) via the low-energy dielectric function of two-dimensional excitons
\begin{equation}
\epsilon_{\mathrm{X}}(\omega) =
\epsilon_{\infty}+\frac{n_{\mathrm{X}} e^2}{\epsilon_0 m}\sum_{q}
\frac{f_{1s,q}}{(\omega^{2}_{1s,q}-\omega^2)-i\omega\Gamma_{\mathrm{X}}}
\end{equation}
\noindent where $\epsilon_{\infty}$ and $\epsilon_{0}$ are the
background and vacuum dielectric constants, respectively, and $e$
and $m$ are the elementary charge and the reduced exciton mass.
The oscillator strengths $f_{1s,q} = (2m\omega_{1s,q}/\hbar)|\langle\psi_{1s}|\textbf{r}|\psi_q\rangle|^2$
for internal transitions from 1s to higher bound and
unbound states are calculated using two-dimensional exciton wavefunctions
$\psi_{i}$ \cite{Hau04,Kai05a}.
We vary the exciton density $n_{\mathrm{X}}$, the energy position
$E_{\mathrm{1s\rightarrow 2p}} = \hbar \omega_{\mathrm{1s,2p}}$, and the
phenomenological exciton broadening parameter $\Gamma_{\mathrm{X}}$ to fit the measured conductivity curves. The
results are shown as dashed curves in Fig.\,2.

At low densities, good agreement with the experimental THz
response is obtained. With increasing pump power, however, the
model does not simultaneously reproduce both $\Delta\sigma_1$ and $\Delta\epsilon_1$. For parameters that optimally fit the measured conductivity (dashed lines in Fig.~2), the corresponding model $\Delta\epsilon_1$ deviates strongly from the experimental result. These problems can be
resolved by considering additional low-frequency spectral weight, in the form of a coexisting plasma of unbound $e$-$h$ pairs. The response of this two-component system is obtained by adding a Drude term to the dielectric function $\epsilon_{\mathrm{X}}$ of Eq.\,(1):
\begin{equation}
\epsilon(\omega) = \epsilon_{\mathrm{X}}(\omega) - \frac{n_{eh}
e^2}{\epsilon_0 m}\frac{1}{\omega^2+i\omega\Gamma_{D}}
\end{equation}
\noindent where $n_{eh}$ is the density of unbound $e$-$h$ pairs, and
$\Gamma_D$ is a phenomenological scattering rate. By including such unbound pairs, excellent agreement with the experiment is obtained even at high
densities (solid curves in Fig. 2). The optimum values
of the parameters are constrained by the different spectral responses of excitons and unbound $e$-$h$ pairs, and by the
requirement to reproduce both functions, $\Delta\sigma_1(\omega)$
and $\Delta\epsilon_1(\omega)$, simultaneously and over a broad
spectral range.

\begin{figure}
\includegraphics[width=8cm]{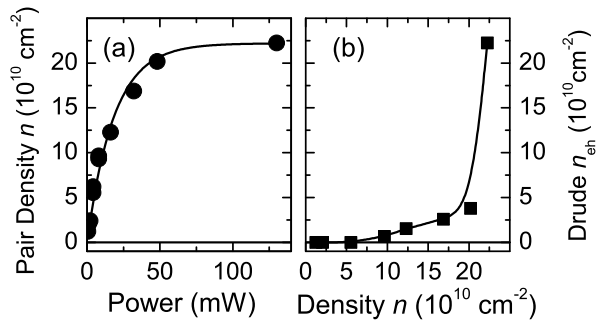}% Here is how to import EPS art
\caption{\label{fig:epsart} Pair densities from the two-component model explained in the text. (a) Total excited pair density as a function of pump power. (b) Free carrier density as a function of the total pair density. Lines are a guide to the eye.}
\end{figure}

\begin{figure}
\includegraphics[width=4.5cm]{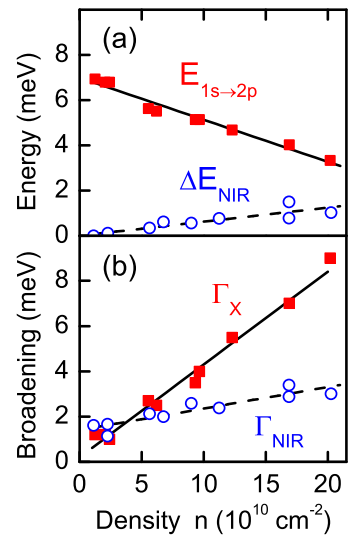}% Here is how to import EPS art
\caption{\label{fig:epsart} (Color online).
(a) Excitonic level spacing $E_{1s\rightarrow 2p}$ (filled squares) measured via THz spectroscopy, and density-induced energy shift $\Delta E_{\mathrm{NIR}}$ (open circles) of the interband resonance. (b) Width $\Gamma_{\mathrm{X}}$ (filled squares) of the intra-excitonic $1s\rightarrow2p$ resonance as fitted via the two-component model of Eq.\,2, corresponding to the solid curves in Fig.\,2. Open circles: width $\Gamma_{\mathrm{NIR}}$ (FWHM) of the near-infrared band edge absorption. Solid and dashed lines: linear fits to the data.}
\end{figure}

The resulting fit parameters reveal details in the evolution of the complex many-body system with increasing excitation density.
Figure~3(a) shows the total density $n = n_{\mathrm{X}}+ n_{eh}$ of excitons and unbound $e$-$h$ pairs as a function of the pump
power. Clearly, a strong saturation at high powers is observed as expected due to bleaching of the near-infrared 1$s$HH line. In the dilute limit, the pair densities can also be estimated from the pump powers (linear near-infrared response). These values agree to within $50\%$ with $n$ obtained from the THz model. In Fig.\,3(b), we plot the density of unbound $e$-$h$ pairs as a function of the total pair excitation $n$. At lower densities ($n < 5\times10^{10}$~cm$^{-2}$), only excitons are generated. With increasing $n$, the population share of unbound carriers becomes more significant. Beyond a critical density $n_C \approx 2 \times 10^{11}$ cm$^{-2}$, photoexcitation results primarily in a population of unbound $e$-$h$ pairs. An excitonic resonance is absent and the THz response can be fully explained by a Drude model alone.

We emphasize that the transition observed here cannot be unambigously identified with a Mott transition driven purely via decreasing inter-particle spacing. Rather, at the highest densities the near-infrared exciton line broadens and shifts, leading to an intrinsic merging with the continuum and thus a simultaneous photoexcitation of unbound carriers. Nevertheless, $n_C$ compares well to values of the Mott density deduced from excitonic and interband photoluminescence after nonresonant excitation~\cite{Kap05}. The THz spectra in Fig.\,2 underscore the utility of low-energy probes to study the crossover between insulating and conducting phases of $e$-$h$ pairs, pointing the way towards THz studies of purely density-driven Mott transition phenomena.

A key aspect of our study concerns the renormalization of the excitonic fine structure with increasing photoexcitation density.  Figure 4 illustrates the changes in resonance energy and scattering rate, via direct quantitative comparison between the THz and the near-infrared response. It shows (filled squares) the energy $E_{1s \rightarrow 2p}$ and width $\Gamma_{\mathrm{X}}$ of the THz exciton transitions. These curves reveal a striking red-shift and broadening that occurs in the 1$s$-2$p$ intra-excitonic oscillator with increasing density. For comparison, the energy shift of the near-infrared 1$s$HH resonance is also indicated [open circles, Fig.\,4(a)]. The net shift is given by the counteracting effects of band gap renormalization and reduced exciton binding energy -- its interpretation is thus elusive and requires model assumptions. In contrast, the intra-excitonic transitions observed here at THz frequencies are unaffected by bandgap shifts, and directly gauge the weakening of the binding energy.

From the near-infrared 1$s$HH absorption band we evaluate the width
$\Gamma_{\mathrm{NIR}}$ as a function of the excitation density [circles in Fig.\,4(b)]. In addition to the spectrometer resolution, the finite optical thickness of the sample leads to a slight inhomogeneous broadening. The measured values of $\Gamma_{\mathrm{NIR}}$ are,
however, a good upper bound for the width of the 1$s$HH
resonance. The density-dependence is well fitted by the relation $\Gamma_{\mathrm{NIR}}(n) = \Gamma_0 + \gamma_{\mathrm{NIR}}n$ [dashed line in Fig.\,4(b)] with $\gamma_{\mathrm{NIR}} = 1.0 \times 10^{-11} {\mathrm{meV cm}^2}$, in quantitative agreement with literature values \cite{Hon89}. Remarkably, the density-dependent broadening of the THz response $\gamma_{\mathrm{X}} = 4.1 \times 10^{-11}
{\mathrm{meV cm}^2}$ (solid line) is about four times larger.
We suggest that this results from enhanced sensitivity of the 2$p$ exciton, not visible in near-infrared spectra, to screening and scattering: its radial extent (expectation value $\langle r_{2p}\rangle = 58\, {\mathrm{nm}}$, obtained from our exciton model in the dilute limit) is six times that of 1$s$ excitons. The combination of near-infrared and THz probes allows for a first direct observation of this enhancement.

In conclusion, we report broadband THz studies of a high-density
exciton gas in GaAs/AlGaAs multiple quantum wells.
With increasing excitation density, the far-infrared complex conductivity of this many-body system reveals a distinct red-shift of the intra-excitonic 1s-2p transition and a strong broadening of the 2p bound state, finally leading to the disappearance of the resonance. THz spectroscopy thus provides a direct gauge of bound and unbound pair densities, and enables the observation of the excitonic fine structure as it evolves under high-density conditions. We believe that this approach will prove valuable in future studies of dense spatially-indirect excitons or interactions of strongly confined carriers in quantum dots.

We thank John Reno (Sandia National Laboratories) for providing quantum-well samples, and Daniel H\"{a}gele, Reinhold L\"{o}venich, and Nils Nielsen for interesting discussions. This work was supported by the Director, Office of Science, Office of Basic Energy Sciences of the US Department of Energy under Contract No. DE-AC02-05CH11231. R.~H. acknowledges support from the Alexander von Humboldt Foundation.

\bibliography{huber_et_al_manuscript}

\begin{thebibliography}{24}
\expandafter\ifx\csname natexlab\endcsname\relax\def\natexlab#1{#1}\fi
\expandafter\ifx\csname bibnamefont\endcsname\relax
  \def\bibnamefont#1{#1}\fi
\expandafter\ifx\csname bibfnamefont\endcsname\relax
  \def\bibfnamefont#1{#1}\fi
\expandafter\ifx\csname citenamefont\endcsname\relax
  \def\citenamefont#1{#1}\fi
\expandafter\ifx\csname url\endcsname\relax
  \def\url#1{\texttt{#1}}\fi
\expandafter\ifx\csname urlprefix\endcsname\relax\def\urlprefix{URL }\fi
\providecommand{\bibinfo}[2]{#2}
\providecommand{\eprint}[2][]{\url{#2}}

\bibitem[{\citenamefont{Rashba}(1982)}]{Ras82}
\bibinfo{author}{\bibfnamefont{E.~I.} \bibnamefont{Rashba}},
  \emph{\bibinfo{title}{Excitons}} (\bibinfo{publisher}{North-Holland Publ.
  Co., Amsterdam}, \bibinfo{year}{1982}).

\bibitem[{\citenamefont{Haug and Koch}(2004)}]{Hau04}
\bibinfo{author}{\bibfnamefont{H.}~\bibnamefont{Haug}} \bibnamefont{and}
  \bibinfo{author}{\bibfnamefont{S.~W.} \bibnamefont{Koch}},
  \emph{\bibinfo{title}{Quantum theory of the Optical and Electronic Properties
  of Semiconductors}} (\bibinfo{publisher}{World Scientific, Signapore},
  \bibinfo{year}{2004}).

\bibitem[{\citenamefont{Keldysh and Kopaev}(1965)}]{Kel65}
\bibinfo{author}{\bibfnamefont{L.~V.} \bibnamefont{Keldysh}} \bibnamefont{and}
  \bibinfo{author}{\bibfnamefont{Y.~V.} \bibnamefont{Kopaev}},
  \bibinfo{journal}{Sov.\ Phys.\ Solid State} \textbf{\bibinfo{volume}{6}},
  \bibinfo{pages}{2219} (\bibinfo{year}{1965}).

\bibitem[{\citenamefont{Mott}(1961)}]{Mot61}
\bibinfo{author}{\bibfnamefont{N.~F.} \bibnamefont{Mott}},
  \bibinfo{journal}{Philos.\ Mag.} \textbf{\bibinfo{volume}{6}},
  \bibinfo{pages}{287} (\bibinfo{year}{1961}).

\bibitem[{\citenamefont{Kappei et~al.}(2005)\citenamefont{Kappei, Szczytko,
  Morier-Genoud, and Deveaud}}]{Kap05}
\bibinfo{author}{\bibfnamefont{L.}~\bibnamefont{Kappei}},
  \bibinfo{author}{\bibfnamefont{J.}~\bibnamefont{Szczytko}},
  \bibinfo{author}{\bibfnamefont{F.}~\bibnamefont{Morier-Genoud}},
  \bibnamefont{and} \bibinfo{author}{\bibfnamefont{B.}~\bibnamefont{Deveaud}},
  \bibinfo{journal}{Phys.\ Rev.\ Lett.} \textbf{\bibinfo{volume}{94}},
  \bibinfo{pages}{147403} (\bibinfo{year}{2005}).

\bibitem[{\citenamefont{Shah}(1999)}]{Sha99}
\bibinfo{author}{\bibfnamefont{J.}~\bibnamefont{Shah}},
  \emph{\bibinfo{title}{Ultrafast Spectroscopy of Semiconductors and
  Semiconductor Nanostructures}} (\bibinfo{publisher}{Springer Verlag},
  \bibinfo{year}{1999}).

\bibitem[{\citenamefont{Chemla and Shah}(2001)}]{Che01}
\bibinfo{author}{\bibfnamefont{D.~S.} \bibnamefont{Chemla}} \bibnamefont{and}
  \bibinfo{author}{\bibfnamefont{J.}~\bibnamefont{Shah}},
  \bibinfo{journal}{Nature} \textbf{\bibinfo{volume}{411}},
  \bibinfo{pages}{549} (\bibinfo{year}{2001}).

\bibitem[{\citenamefont{Honold et~al.}(1989)\citenamefont{Honold, Schultheis,
  Kuhl, and Tu}}]{Hon89}
\bibinfo{author}{\bibfnamefont{A.}~\bibnamefont{Honold}},
  \bibinfo{author}{\bibfnamefont{L.}~\bibnamefont{Schultheis}},
  \bibinfo{author}{\bibfnamefont{J.}~\bibnamefont{Kuhl}}, \bibnamefont{and}
  \bibinfo{author}{\bibfnamefont{C.~W.} \bibnamefont{Tu}},
  \bibinfo{journal}{Phys.\ Rev.\ B} \textbf{\bibinfo{volume}{40}},
  \bibinfo{pages}{6442} (\bibinfo{year}{1989}).

\bibitem[{\citenamefont{Tr{\"{a}}nkle et~al.}(1987)\citenamefont{Tr{\"{a}}nkle,
  Lach, Forchel, Scholz, Ell, Haug, Weimann, Griffiths, Kroemer, and
  Subbanna}}]{Tra87}
\bibinfo{author}{\bibfnamefont{G.}~\bibnamefont{Tr{\"{a}}nkle}},
  \bibinfo{author}{\bibfnamefont{E.}~\bibnamefont{Lach}},
  \bibinfo{author}{\bibfnamefont{A.}~\bibnamefont{Forchel}},
  \bibinfo{author}{\bibfnamefont{F.}~\bibnamefont{Scholz}},
  \bibinfo{author}{\bibfnamefont{C.}~\bibnamefont{Ell}},
  \bibinfo{author}{\bibfnamefont{H.}~\bibnamefont{Haug}},
  \bibinfo{author}{\bibfnamefont{G.}~\bibnamefont{Weimann}},
  \bibinfo{author}{\bibfnamefont{G.}~\bibnamefont{Griffiths}},
  \bibinfo{author}{\bibfnamefont{H.}~\bibnamefont{Kroemer}}, \bibnamefont{and}
  \bibinfo{author}{\bibfnamefont{S.}~\bibnamefont{Subbanna}},
  \bibinfo{journal}{Phys.\ Rev.\ B} \textbf{\bibinfo{volume}{36}},
  \bibinfo{pages}{6712} (\bibinfo{year}{1987}).

\bibitem[{\citenamefont{Choi et~al.}(2004)\citenamefont{Choi, Je, Yim, and
  Park}}]{Cho04}
\bibinfo{author}{\bibfnamefont{M.}~\bibnamefont{Choi}},
  \bibinfo{author}{\bibfnamefont{K.-C.} \bibnamefont{Je}},
  \bibinfo{author}{\bibfnamefont{S.-Y.} \bibnamefont{Yim}}, \bibnamefont{and}
  \bibinfo{author}{\bibfnamefont{S.-H.} \bibnamefont{Park}},
  \bibinfo{journal}{Phys.\ Rev.\ B} \textbf{\bibinfo{volume}{70}},
  \bibinfo{pages}{085309} (\bibinfo{year}{2004}).

\bibitem[{\citenamefont{Fehrenbach et~al.}(1982)\citenamefont{Fehrenbach,
  Sch{\"{a}}fer, Treusch, and Ulbrich}}]{Feh82}
\bibinfo{author}{\bibfnamefont{G.~W.} \bibnamefont{Fehrenbach}},
  \bibinfo{author}{\bibfnamefont{W.}~\bibnamefont{Sch{\"{a}}fer}},
  \bibinfo{author}{\bibfnamefont{J.}~\bibnamefont{Treusch}}, \bibnamefont{and}
  \bibinfo{author}{\bibfnamefont{R.~G.} \bibnamefont{Ulbrich}},
  \bibinfo{journal}{Phys.\ Rev.\ Lett.} \textbf{\bibinfo{volume}{49}},
  \bibinfo{pages}{1281} (\bibinfo{year}{1982}).

\bibitem[{\citenamefont{Peyghambarian et~al.}(1984)\citenamefont{Peyghambarian,
  Gibbs, Jewell, Antonetti, Migus, Hulin, and Mysyrowicz}}]{Pey84}
\bibinfo{author}{\bibfnamefont{N.}~\bibnamefont{Peyghambarian}},
  \bibinfo{author}{\bibfnamefont{H.~M.} \bibnamefont{Gibbs}},
  \bibinfo{author}{\bibfnamefont{J.~L.} \bibnamefont{Jewell}},
  \bibinfo{author}{\bibfnamefont{A.}~\bibnamefont{Antonetti}},
  \bibinfo{author}{\bibfnamefont{A.}~\bibnamefont{Migus}},
  \bibinfo{author}{\bibfnamefont{D.}~\bibnamefont{Hulin}}, \bibnamefont{and}
  \bibinfo{author}{\bibfnamefont{A.}~\bibnamefont{Mysyrowicz}},
  \bibinfo{journal}{Phys.\ Rev.\ Lett.} \textbf{\bibinfo{volume}{53}},
  \bibinfo{pages}{2433} (\bibinfo{year}{1984}).

\bibitem[{\citenamefont{Wake et~al.}(1992)\citenamefont{Wake, Yoon, Wolfe, and
  Morkoc}}]{Wak92}
\bibinfo{author}{\bibfnamefont{D.~R.} \bibnamefont{Wake}},
  \bibinfo{author}{\bibfnamefont{H.~W.} \bibnamefont{Yoon}},
  \bibinfo{author}{\bibfnamefont{J.~P.} \bibnamefont{Wolfe}}, \bibnamefont{and}
  \bibinfo{author}{\bibfnamefont{H.}~\bibnamefont{Morkoc}},
  \bibinfo{journal}{Phys.\ Rev.\ B} \textbf{\bibinfo{volume}{46}},
  \bibinfo{pages}{13452} (\bibinfo{year}{1992}).

\bibitem[{\citenamefont{Manzke et~al.}(1998)\citenamefont{Manzke, Peng,
  Henneberger, Neukirch, Hauke, Wundke, Gutowski, and Hommel}}]{Man98}
\bibinfo{author}{\bibfnamefont{G.}~\bibnamefont{Manzke}},
  \bibinfo{author}{\bibfnamefont{Q.~Y.} \bibnamefont{Peng}},
  \bibinfo{author}{\bibfnamefont{K.}~\bibnamefont{Henneberger}},
  \bibinfo{author}{\bibfnamefont{U.}~\bibnamefont{Neukirch}},
  \bibinfo{author}{\bibfnamefont{K.}~\bibnamefont{Hauke}},
  \bibinfo{author}{\bibfnamefont{K.}~\bibnamefont{Wundke}},
  \bibinfo{author}{\bibfnamefont{J.}~\bibnamefont{Gutowski}}, \bibnamefont{and}
  \bibinfo{author}{\bibfnamefont{D.}~\bibnamefont{Hommel}},
  \bibinfo{journal}{Phys.\ Rev.\ Lett.} \textbf{\bibinfo{volume}{80}},
  \bibinfo{pages}{4943} (\bibinfo{year}{1998}).

\bibitem[{\citenamefont{Litvinenko et~al.}(1999)\citenamefont{Litvinenko,
  Birkedahl, Lyssenko, and Hvam}}]{Lit99}
\bibinfo{author}{\bibfnamefont{K.}~\bibnamefont{Litvinenko}},
  \bibinfo{author}{\bibfnamefont{D.}~\bibnamefont{Birkedahl}},
  \bibinfo{author}{\bibfnamefont{V.~G.} \bibnamefont{Lyssenko}},
  \bibnamefont{and} \bibinfo{author}{\bibfnamefont{J.~M.} \bibnamefont{Hvam}},
  \bibinfo{journal}{Phys.\ Rev.\ B} \textbf{\bibinfo{volume}{59}},
  \bibinfo{pages}{10255} (\bibinfo{year}{1999}).

\bibitem[{\citenamefont{Knox et~al.}(1986)\citenamefont{Knox, Hirlimann,
  Miller, Shah, Chemla, and Shank}}]{Kno86}
\bibinfo{author}{\bibfnamefont{W.~H.} \bibnamefont{Knox}},
  \bibinfo{author}{\bibfnamefont{C.}~\bibnamefont{Hirlimann}},
  \bibinfo{author}{\bibfnamefont{D.~A.~B.} \bibnamefont{Miller}},
  \bibinfo{author}{\bibfnamefont{J.}~\bibnamefont{Shah}},
  \bibinfo{author}{\bibfnamefont{D.~S.} \bibnamefont{Chemla}},
  \bibnamefont{and} \bibinfo{author}{\bibfnamefont{C.~V.} \bibnamefont{Shank}},
  \bibinfo{journal}{Phys.\ Rev.\ Lett.} \textbf{\bibinfo{volume}{56}},
  \bibinfo{pages}{1191} (\bibinfo{year}{1986}).

\bibitem[{\citenamefont{Timusk}(1976)}]{Tim76}
\bibinfo{author}{\bibfnamefont{T.}~\bibnamefont{Timusk}},
  \bibinfo{journal}{Phys.\ Rev.\ B} \textbf{\bibinfo{volume}{13}},
  \bibinfo{pages}{3511} (\bibinfo{year}{1976}).

\bibitem[{\citenamefont{Groeneveld and Grischkowsky}(1994)}]{Gro94}
\bibinfo{author}{\bibfnamefont{R.~H.~M.} \bibnamefont{Groeneveld}}
  \bibnamefont{and}
  \bibinfo{author}{\bibfnamefont{D.}~\bibnamefont{Grischkowsky}},
  \bibinfo{journal}{J.\ Opt.\ Soc.\ Am.\ B} \textbf{\bibinfo{volume}{11}},
  \bibinfo{pages}{2502} (\bibinfo{year}{1994}).

\bibitem[{\citenamefont{{\v{C}}erne et~al.}(1996)\citenamefont{{\v{C}}erne,
  Kono, Sherwin, Sundaram, Gossard, and Bauer}}]{Cer96}
\bibinfo{author}{\bibfnamefont{J.}~\bibnamefont{{\v{C}}erne}},
  \bibinfo{author}{\bibfnamefont{J.}~\bibnamefont{Kono}},
  \bibinfo{author}{\bibfnamefont{M.~S.} \bibnamefont{Sherwin}},
  \bibinfo{author}{\bibfnamefont{M.}~\bibnamefont{Sundaram}},
  \bibinfo{author}{\bibfnamefont{A.~C.} \bibnamefont{Gossard}},
  \bibnamefont{and} \bibinfo{author}{\bibfnamefont{G.~E.~W.}
  \bibnamefont{Bauer}}, \bibinfo{journal}{Phys.\ Rev.\ Lett.}
  \textbf{\bibinfo{volume}{77}}, \bibinfo{pages}{1131} (\bibinfo{year}{1996}).

\bibitem[{\citenamefont{Kira et~al.}(2000)\citenamefont{Kira, Hoyer, Stroucken,
  and Koch}}]{Kir00}
\bibinfo{author}{\bibfnamefont{M.}~\bibnamefont{Kira}},
  \bibinfo{author}{\bibfnamefont{W.}~\bibnamefont{Hoyer}},
  \bibinfo{author}{\bibfnamefont{T.}~\bibnamefont{Stroucken}},
  \bibnamefont{and} \bibinfo{author}{\bibfnamefont{S.~W.} \bibnamefont{Koch}},
  \bibinfo{journal}{Phys.\ Rev.\ Lett.} \textbf{\bibinfo{volume}{87}},
  \bibinfo{pages}{176401} (\bibinfo{year}{2000}).

\bibitem[{\citenamefont{Kaindl et~al.}(2003)\citenamefont{Kaindl, Carnahan,
  H{\"{a}}gele, L{\"{o}}venich, and Chemla}}]{Kai03}
\bibinfo{author}{\bibfnamefont{R.~A.} \bibnamefont{Kaindl}},
  \bibinfo{author}{\bibfnamefont{M.~A.} \bibnamefont{Carnahan}},
  \bibinfo{author}{\bibfnamefont{D.}~\bibnamefont{H{\"{a}}gele}},
  \bibinfo{author}{\bibfnamefont{R.}~\bibnamefont{L{\"{o}}venich}},
  \bibnamefont{and} \bibinfo{author}{\bibfnamefont{D.~S.}
  \bibnamefont{Chemla}}, \bibinfo{journal}{Nature}
  \textbf{\bibinfo{volume}{423}}, \bibinfo{pages}{734} (\bibinfo{year}{2003}).

\bibitem[{Gal()}]{Gal05}
\bibinfo{note}{I. Galbraith {\it et al.}, Phys. Rev. B {\bf 71}, 073302
  (2005).}

\bibitem[{Hub()}]{Hub01}
\bibinfo{note}{R. Huber, F. Tauser, A. Brodschelm, M. Bichler, G. Abstreiter,
  and A. Leitenstorfer, Nature {\bf 414}, 286 (2001); R. Huber, C.
  K{\"{u}}bler, S. T{\"{u}}bel, A. Leitenstorfer, Q. T. Vu, H. Haug, F.
  K{\"{o}}hler, and M.-C. Amann, Phys. Rev. Lett. {\bf 94}, 027401 (2005).}

\bibitem[{Kai()}]{Kai05a}
\bibinfo{note}{R. A. Kaindl, D. H{\"{a}}gele, M. A. Carnahan, R.
  L{\"{o}}venich, and D. S. Chemla (to be published)}.

\end{thebibliography}
\end{document}